\documentclass[pra,aps,amsmath,amssymb,twocolumn,floatfix,superscriptaddress]{revtex4-2}
\usepackage{amsmath}
\usepackage{graphicx}
\usepackage{color}
\usepackage{amssymb}
\usepackage[utf8]{inputenc}
\usepackage[T1]{fontenc}

\usepackage{comment}
\usepackage[normalem]{ulem}
\usepackage{hyperref}
\hypersetup{colorlinks,
 linkcolor=blue,%
 citecolor=blue,%
 urlcolor=blue
}
\usepackage{array}

\begin{document}
\title{Valley polarization dependence of candidate even-denominator fractional quantum Hall states in monolayer graphene}
\author{Saswata Sahu }
\email{sahu.saswata@gmail.com}\affiliation{Department of Basic Science \& Humanities, Abacus Institute of Engineering and Management, Hooghly - 712148, India} 
\author{Amarendra Pr. Indra }
\email{indraamarendra@gmail.com}\affiliation{Sutragarh M.N. High School, Santipur, Nadia -741404, India} 
\author{Moumita Indra } 
\email{moumita.indra@saha.ac.in}
\affiliation{Theory Division, Saha Institute of Nuclear Physics, 1/AF Bidhannagar, Kolkata - 700064, India}
\date{\today}
\begin{abstract}
We investigate even-denominator fractional quantum Hall (EDFQH) states in monolayer graphene, focusing on the experimentally observed filling fractions $\nu = 1/2$ and $\nu = 1/4$. Owing to the approximate SU(4) symmetry arising from spin and valley degrees of freedom, graphene hosts a rich variety of multi-component fractional quantum Hall states with different polarization configurations. Using the Chern–Simons composite fermion framework for SU(4) systems, we construct candidate variational wave functions corresponding to distinct spin, valley, and mixed polarization states. The ground-state energies of these states are evaluated numerically using the Coulomb interaction in spherical geometry. By comparing energies across different polarization sectors, we identify the energetically favored configurations at $\nu = 1/2$ and $\nu = 1/4$. 
Our variational energy comparison suggests that the energetically favored candidate states correspond to multicomponent correlated quantum Hall liquids. Within the restricted family of trial states considered here, these states differ from a simple composite-fermion Fermi-sea description.
These findings provide insight into the role of valley polarization and SU(4) symmetry in stabilizing even-denominator fractional quantum Hall states in graphene.
\end{abstract}
\maketitle 
\section{Introduction}
The fractional quantum Hall (FQH) effect is a paradigmatic example of a strongly correlated quantum phase emerging in two-dimensional electron systems subjected to a strong perpendicular magnetic field at low temperatures. In this regime, the kinetic energy of electrons is quenched into highly degenerate Landau levels, and electron–electron interactions drive the formation of incompressible quantum liquids characterized by fractionally quantized Hall conductance \cite{Tsui_PRL:1982, Tsui_PRB:1982}. While the majority of experimentally observed FQH states occur at odd-denominator filling fractions and are well described within the composite fermion framework, even-denominator fractional quantum Hall (EDFQH) states \cite{Shabani:2009, Wang:2022, Wang:2025} remain comparatively rare and continue to attract significant theoretical and experimental interest due to their unconventional nature \cite{Hossain:2018, sha:2025} and potential connection to non-Abelian quasi-particles \cite{Nayak:2008, Huang:2026}.

Monolayer graphene \cite{Ajit:2015} provides a unique platform for exploring FQH physics beyond conventional semiconductor hetero-structures. Owing to its honeycomb lattice structure, graphene hosts two in-equivalent valleys, labeled $K$ and $K'$, in addition to the spin degree of freedom, giving rise to an approximate SU(4) internal symmetry in the absence of symmetry-breaking perturbations \cite{Geim:2009, MacDonald:2006, Shibata:2008}. As a result, each Landau level in graphene is fourfold degenerate, allowing for a rich landscape of multi-component FQH states with distinct spin and valley polarization configurations \cite{Dean:2011, Indra:2024_3}. Advances in sample quality and device engineering have enabled the experimental observation of a variety of symmetry-broken and symmetry-preserving FQH states in graphene, including EDFQH states at filling fractions such as $\nu = 1/2$ and $\nu = 1/4$ \cite{Li:2017}.

The nature of EDFQH states in graphene \cite{Papic:2011} is fundamentally different from their counterparts in conventional two-dimensional electron gases \cite{Eisenstein:1994}. In particular, the interplay between Coulomb interactions, Zeeman coupling, and valley anisotropies determines whether the system favors spin-polarized, valley-polarized, or mixed-polarization ground states \cite{Modak:2011}. Experimental studies \cite{Young:2012, MacDonald:2014, Kharitonov:2012} have highlighted a pronounced contrast between spin and valley degrees of freedom in graphene, as inter-valley scattering requires large momentum transfer, whereas spin-flip processes involve relatively small energy scales. This distinction suggests that valley polarization can play a dominant role in stabilizing certain FQH states, motivating theoretical investigations \cite{Sujit:2018} that explicitly incorporate the SU(4) structure of graphene.

Several theoretical approaches have been proposed to describe EDFQH states in monolayer graphene. Multi-component composite fermion \cite{Balram:2015} and parton constructions \cite{Kim:2019} have been employed to capture the role of internal symmetries, while numerical studies have explored candidate incompressible states and their collective excitations. Despite this progress, the polarization structure and energetic stability of EDFQH states at $\nu = 1/2$ and $\nu = 1/4$ within an SU(4) framework remain open questions, particularly in the context of realistic Coulomb interactions.

In this work, we address these issues by systematically analyzing candidate EDFQH states in monolayer graphene using the Chern–Simons composite fermion theory \cite{Lopez:1991} for SU(4) systems. We construct variational wave functions corresponding to different combinations of spin, valley, and mixed polarizations and evaluate their ground-state energies using the Coulomb interaction in spherical geometry. By comparing energies across polarization sectors at $\nu = 1/2$ and $\nu = 1/4$, we identify the energetically favored states and examine the role of valley polarization in candidate EDFQH states
stabilizing incompressible EDFQH phases. 
Our results identify energetically competitive multi-component candidate states within the SU(4) Chern–Simons framework, highlighting the importance of multi-component correlations in graphene-based FQH systems.

In this work, we revisit even-denominator fractional quantum Hall states in monolayer graphene within the SU(4) Chern–Simons composite-fermion framework, with particular emphasis on the role of valley polarization. Unlike Ref. \cite{Indra:2024_3}, which examined polarization effects more broadly, the present work focuses specifically on the dependence of Coulomb energetics on valley polarization and provides a systematic comparison among multiple valley-polarization sectors. We perform a systematic variational comparison of candidate states across multiple valley-polarization sectors at ($\nu = 1/2$), including partially and fully polarized states. For ($\nu = 1/4$), where a larger degeneracy of candidate configurations arises, we examine competing fully valley-polarized sectors with different flux-attachment classes and analyze their energetic ordering. By comparing Coulomb ground-state energies in spherical geometry, we identify energetically competitive candidate states and discuss their relevance to experimentally observed even-denominator fillings in monolayer graphene.

The remainder of the paper is organized as follows. In Sec. II, we introduce the SU(4) Chern–Simons formalism and define the polarization indices and flux-attachment scheme. The construction of trial wave functions and the details of the energy calculations are presented in Sec. III. Our numerical results and their physical implications are discussed in Sec. IV, followed by conclusions in Sec. V.
\section{MODEL AND FORMULATION}
We study even-denominator fractional quantum Hall (EDFQH) states in monolayer graphene by employing the Chern–Simons (CS) composite fermion theory for an SU(4) system. In graphene, the four-fold degeneracy of each Landau level arises from the combination of spin $(\uparrow, \downarrow)$ and valley $(K, K')$, degrees of freedom. We denote these four components by $\alpha=1,2,3,4$, corresponding to $(+ \uparrow ), (+ \downarrow ), (- \uparrow ), (- \downarrow )$ respectively.

\subsection{Chern–Simons flux attachment}
Within the CS approach, each quasi-particle of species $\alpha$ experiences an effective magnetic field given by
\begin{equation}
 B^*_\alpha = B - \phi_0 \; {\cal{K}}_{\alpha \beta}  \; \rho_ \beta 
\end{equation}
where $B$ is the applied actual magnetic field, $\phi_0$ is the flux quanta and $\rho_\beta$ represents density of the electrons for the species $\beta$.
${\cal K}_{\alpha \beta}$ is a symmetric $4 \times 4$ matrix encoding the flux attachment between different components. We adopt a simplified flux attachment scheme, consistent with previous studies, where
\begin{equation}
\cal{K} = \left(
      \begin{array}{llll}
         2k_1 & m_1 &n_1 & n_2 \\
         m_1 & 2k_2 & n_3 & n_4 \\
         n_1 & n_2  & 2k_3 & m_2 \\
         n_3 & n_4 & m_2 & 2k_4
      \end{array}
\right), \label{K-matrix}
\end{equation}
\begin{figure}
 \centering
  {\includegraphics[width=0.5\textwidth]{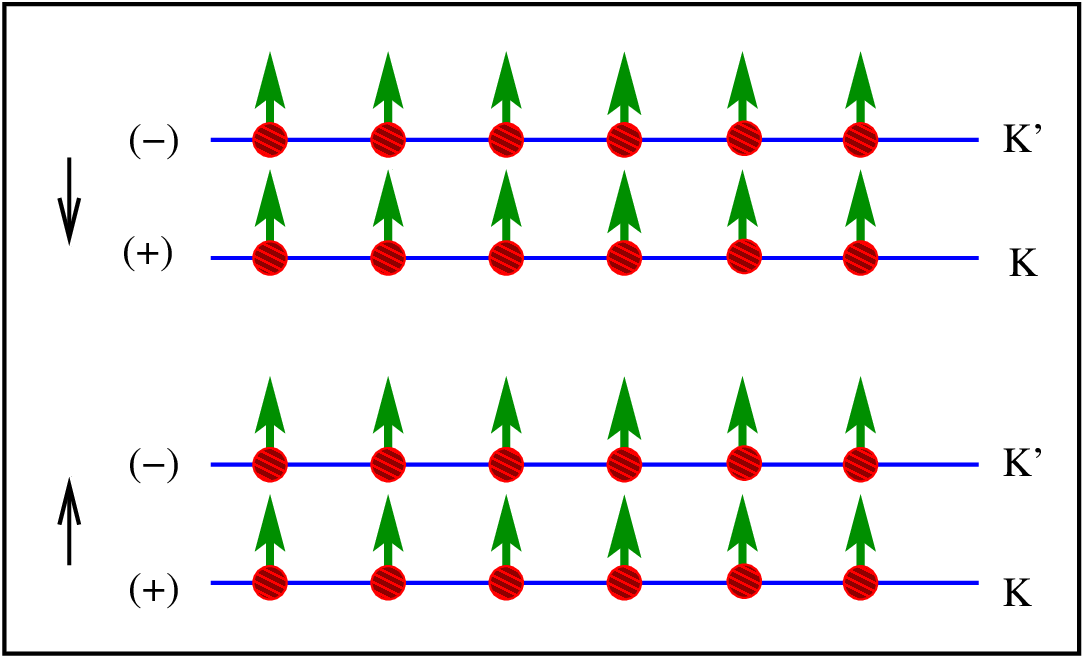}}
   \caption{Four states in monolayer Graphene: Two spin ($\uparrow$ and $\downarrow$) states and two valley ($K$ and $K'$) states represented as $+$ and $-$ valley}
   \label{Fig1}
 \end{figure}

For simplicity, we set $k_1 = k_2, k_3=k_4$, and all off-diagonal elements $n_i=n$. The filling factors of the CS quasiparticles, $\nu_{\alpha}$, are related to the electron filling factor $\nu$ by
\begin{equation}
\rho_\alpha / \nu_\alpha = \rho / \nu - {\cal K}_{\alpha\beta} \rho_\beta \;.
\label{eff}
\end{equation}
where, $\rho=\sum_{\alpha} \rho_{\alpha}$ is the total electron density.
\subsection{Polarization indices}
To characterize the spin and valley structure of a given FQH state, we define the spin (S), the valley (V), and the mixed (M) polarization indices:
 \begin{eqnarray}
 S = (\rho_1 + \rho_3 - \rho_2 - \rho_4)/ \rho,  \nonumber \\
 V = (\rho_1 + \rho_2 - \rho_3 -\rho_4 )/ \rho, \nonumber \\
 M = (\rho_1 + \rho_4 - \rho_2 - \rho_3) / \rho, 
 \end{eqnarray}
These indices satisfy $|S|,|V|,|M|\le 1$ and capture the relative population of the four SU(4) components. The choice of flux attachment parameters $(k_1, k_3, m_1, m_2, n)$ along with 
$(S,V,M)$ uniquely determines the electron filling fraction $\nu$ for the state.

\subsection{Spherical geometry}
For numerical calculations, we consider $N=N_1+N_2+N_3+N_4$ electrons confined to the surface of a sphere, pierced by a radial magnetic field generated by a Dirac monopole of strength $Q$ \cite{Haldane:1983, Jain:1989}. The effective monopole strength for species $\alpha$ is
\begin{equation}
  2q_\alpha = 2Q -  \sum_{\beta} (N_\beta-\delta_{\alpha \beta})\cal{K}_{\alpha \beta}
\end{equation}
\begin{eqnarray*}
  \sum_{\alpha} 2 q_\alpha &=& 8 Q - \sum_{\alpha} \sum_{\beta} \left ( N_\alpha -\delta_{\alpha \beta} \right ) \cal{K}_{\alpha \beta}\\
\Rightarrow Q &=& \frac{1}{8} \left (\sum_{\alpha} 2q_\alpha + \sum_{\alpha} {\cal K}_{\alpha\alpha} (N_\alpha-1) \right) \\
           &+&\frac{1}{8} \left(m_1 N_1+m_1 N_2+m_2 N_3 +m_2 N_4+ 2nN \right)
\end{eqnarray*}
2$Q\phi_0$ is the total flux through the spherical surface of radius $R$, if $B$ is the corresponding magnetic field then
\begin{eqnarray*}
  2Q \phi_0 = 4\pi R^2 B
  \Rightarrow R^2 = \frac{2Q}{4\pi} \frac{hc}{e} \frac{e l^2}{\hbar c}  = Q {l_0}^2
\end{eqnarray*}
So the radius of the sphere is $R = \sqrt{Q}$ in unit of magnetic length $l_0$.
We have chosen the values of $N_1$, $N_2$, $N_3$, $N_4$ such a way that it will fix the values of $S$, $V$, $M$.
\subsection{Trial Wave-function}
Let us suppose that there are $\nu_1$ number of spin-up $(+)$ valley $\Lambda$-level, $\nu_2$ number of spin-down $(+)$ valley
$\Lambda$-level, $\nu_3$ number of spin-up $(-)$ valley $\Lambda$-level, $\nu_4$ number of spin-down $(-)$ valley $\Lambda$-level. 
Using Eq. (\ref{eff}), the total filling factor $\nu$ can be expressed in terms of the polarization indices $(S,V,M)$, the flux-attachment parameters ($k_1, k_3, m_1, m_2, n$)  and the composite-fermion filling factors ($\nu_1, \nu_2, \nu_3, \nu_4$).


\begin{subequations}\label{eq:filling}
\begin{align}
&(1+V+S+M)\left(2k_1+\frac{1}{\nu_1}\right)
 +(1+V-S-M)m_1 \nonumber\\
&\qquad +2n(1-V)=\frac{4}{\nu}, \label{eq:6a}\\[-1mm]
&(1+V-S-M)\left(2k_1+\frac{1}{\nu_2}\right)
 +(1+V+S+M)m_1 \nonumber\\
&\qquad +2n(1-V)=\frac{4}{\nu}, \label{eq:6b}\\[-1mm]
&(1-V+S+M)\left(2k_3+\frac{1}{\nu_3}\right)
 +(1-V-S+M)m_2 \nonumber\\
&\qquad +2n(1+V)=\frac{4}{\nu}, \label{eq:6c}\\[-1mm]
&(1-V-S+M)\left(2k_3+\frac{1}{\nu_4}\right)
 +(1-V+S-M)m_2 \nonumber\\
&\qquad +2n(1+V)=\frac{4}{\nu}. \label{eq:6d}
\end{align}
\label{eqn6}
\end{subequations}

The five indices $k_1$, $k_3$, $m_1$, $m_2$, $n$ mimic the interaction strength between different CF $\Lambda$-levels. 
Equations (\ref{eq:6a})–(\ref{eq:6d}) show that the total filling fraction $\nu$ is uniquely determined by the polarization indices $(S,V,M)$, the composite-fermion filling factors ($\nu_1, \nu_2, \nu_3, \nu_4$), and the interaction parameters ($k_1, k_3, m_1, m_2, n$) and the particular state is symbolized as ($k_1 k_3 m_1 m_2 n$) following the work \cite{Indra:2024_3}.

\begin{figure*}
 \centering
    {\includegraphics[width=0.625\textwidth]{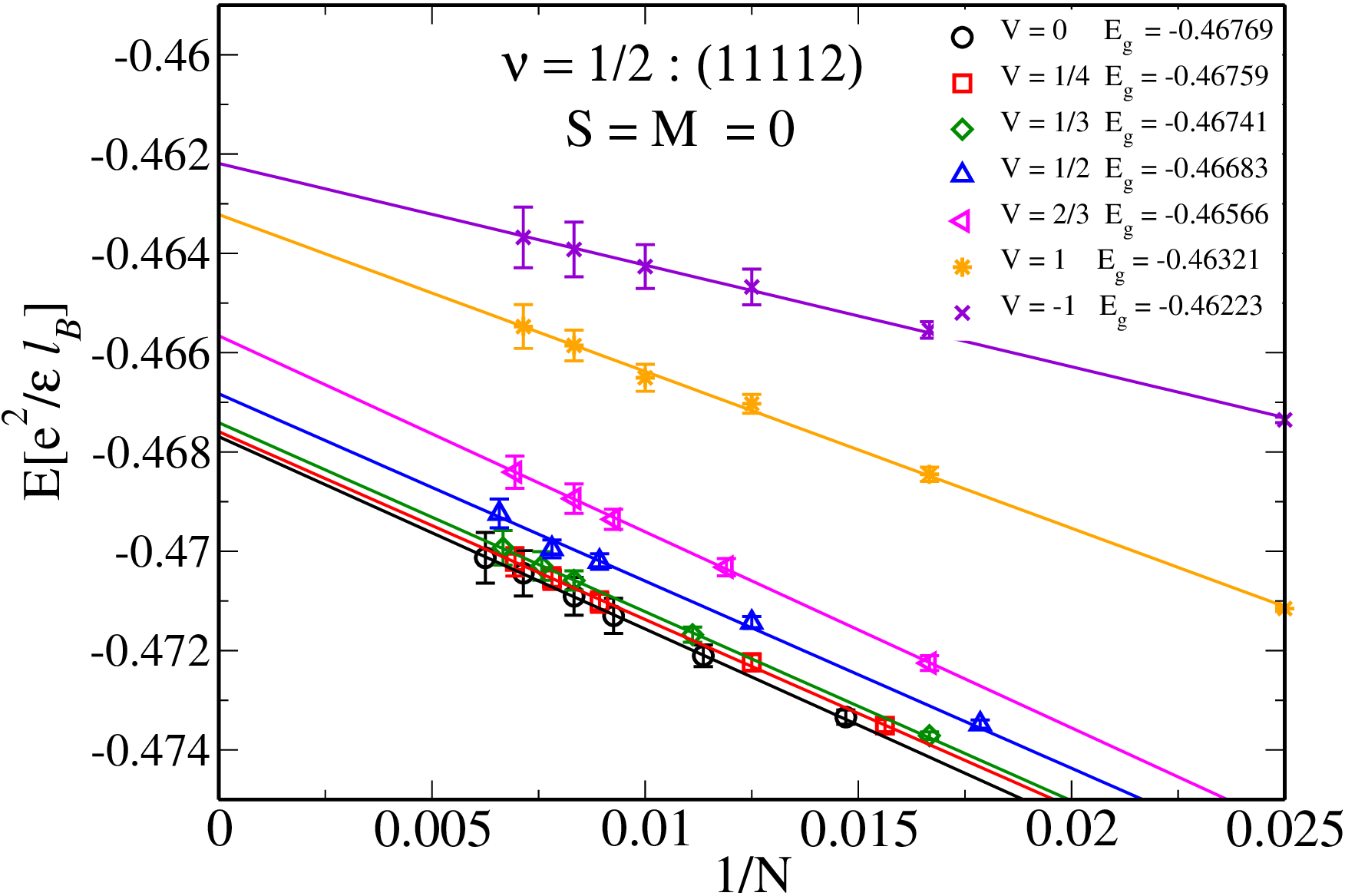}}
   \caption{Ground-state energy per particle as a function of the inverse particle number $1/N$ for different valley polarizations $V$, with $S=M=0$, at filling fraction $\nu=1/2$. The flux-attachment parameter set $(k_1 k_3 m_1 m_2 n)$ is fixed in this case to ( 1 1 1 1 2). Error bars represent the Monte Carlo statistical uncertainties of the finite-size energies; for most data points, they are smaller than the marker size. The thermodynamic limit corresponds to $1/N \rightarrow 0$.}
   \label{Fig2}
   
   {\includegraphics[width=0.625\textwidth]{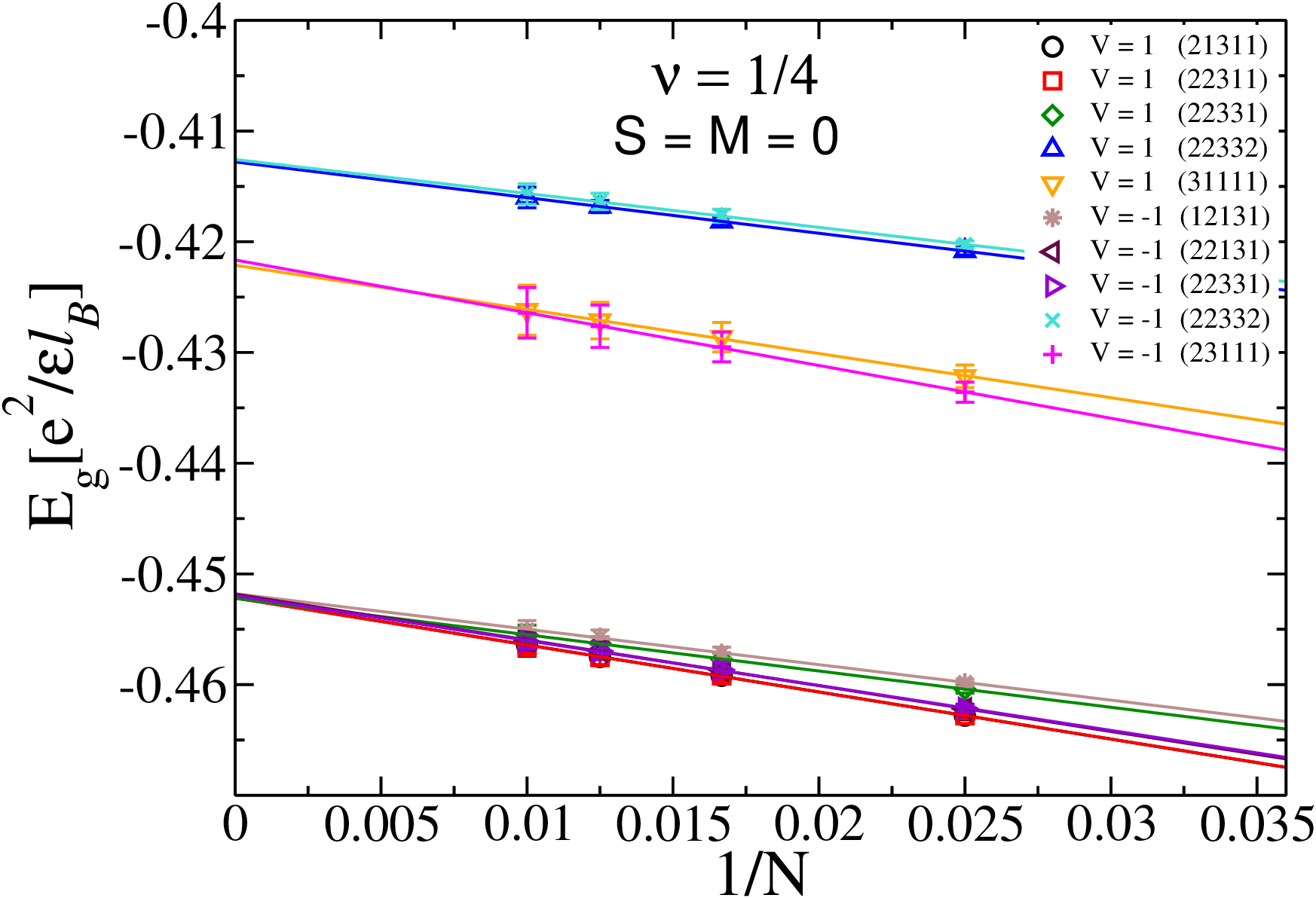}}
   \caption{Ground-state energy per particle for different flux-attachment parameter sets $(k_1 k_3 m_1 m_2 n)$ in the fully valley-polarized sectors $(V=\pm1)$, with $S=M=0$, at filling fraction $\nu=1/4$. Error bars denote the Monte Carlo statistical uncertainties and are smaller than the marker size for most data points.}
   
   \label{Fig3}
 \end{figure*}

Following Modak \textit{et al.} \cite{Modak:2011}, the variational wave function for a multi-component FQH state is constructed as
\begin{eqnarray}
\Psi _{\{\kappa_{\alpha \beta }\}}  &=&  {\cal P}_L \Phi_{\nu_1}(\Omega_1^1,\cdots,\Omega_{N_1}^1) \Phi_{\nu_2}(\Omega_1^2,\cdots,\Omega_{N_2}^2)  \nonumber \\
&&\times \Phi_{\nu_3}(\Omega_1^3,\cdots,\Omega_{N_3}^3) \Phi_{\nu_4}(\Omega_1^4,\cdots,\Omega_{N_4}^4) 
\nonumber \\
&&\times J_{11} \;J_{22}\; J_{33}\; J_{44}\; J_{12} \;J_{34}\; J_{13}\; J_{14}\; J_{23}\; J_{24} \nonumber 
\label{Wavefn}
\end{eqnarray}
where, $\Phi_{\nu_1}$ is the Slater determinant of $\nu_1$ filled $\Lambda$ level CFs, $\Omega_1, \cdots \Omega_{N_{\alpha}}$ are the positions of CFs on the spherical surface, upper index indicate the different species of CFs and the Jastrow factor is given by
\begin{eqnarray}
  J_{\alpha \beta} &=&  \prod_{i, j}^{N_\alpha, N_\beta} (u_i^{(\alpha)} v_j^{(\beta)}-u_j^{(\beta)} v_i^{(\alpha)})^{\kappa_{\alpha \beta }} \mbox{~~~~ if $\alpha \; \neq \beta$ }\nonumber \\
  J_{\alpha \alpha}& =&  \prod_{i< j}^{N_\alpha} (u_i^{(\alpha)} v_j^{(\alpha)}-u_j^{(\alpha)} v_i^{(\alpha)})^{\kappa_{\alpha \alpha }} \nonumber 
  \end{eqnarray}
The prefix within bracket represent the CFs of different $\Lambda$ levels. Here we have used the spinor coordinates 
\begin{eqnarray*}
  u(\Omega) = cos(\theta/2) e^{-i\phi/2} \mbox{~~~ and ~~~~} v(\Omega) = sin (\theta/2) e^{i\phi/2}
\end{eqnarray*}
Some of the above wave functions are closely related to the Halperin-type multi-component states derived within the plasma analogy framework \cite{Halperin:1984, Goerbig:2008}. 
The ground-state energy per particle is computed using the Coulomb interaction,
\begin{eqnarray}
  E_g 
  &=& \left [ \frac{<\Psi _{\{\kappa_{\alpha \beta}\}} | H | \Psi _{\{\kappa_{\alpha \beta}\}}> }{<\Psi _{\{\kappa_{\alpha \beta}\}}| \Psi _{\{\kappa_{\alpha \beta}\}}>} - \frac{N^2 e^2}{2 \epsilon R} \right] /N \;
\end{eqnarray}
Here, $H = \sum_{i<j} \frac{e^2}{r_{i,j}}$ is the Coulomb interaction, with $r_{i,j}$ as the inter-electronic distance and  and the second term accounts for the neutralizing background.
 In the present work, Zeeman energy is neglected and the calculations are performed within a Coulomb-only SU(4)-symmetric framework. The Coulomb ground-state energies were evaluated numerically using Monte Carlo sampling with more than ($5 \times 10^5$) Monte Carlo steps for each system size. Calculations were performed for systems containing up to ($N=120$) particles. The thermodynamic-limit energies were obtained by linearly extrapolating the finite-size energies as a function of ($1/N$) using a best-fit procedure. The statistical uncertainties obtained from the Monte Carlo sampling are shown as error bars in Fig. \ref{Fig2} and Fig. \ref{Fig3}. For most system sizes, these uncertainties are smaller than the marker size.

\section{RESULTS AND INTERPRETATION}
We now discuss the numerical results obtained for the ground-state energies of candidate even-denominator fractional quantum Hall states at filling fractions $\nu = 1/2$ and $\nu = 1/4$
in monolayer graphene. For each filling fraction, we evaluate the Coulomb ground-state energy per particle for different choices of flux-attachment parameters $(k_1 k_3 m_1 m_2n)$, corresponding to distinct spin, valley, and mixed polarization configurations. The calculations are performed in spherical geometry for finite system sizes, and the energies are extrapolated to the thermodynamic limit $N \rightarrow \infty$.

\subsection{Filling fraction $\nu = 1/2$}
Figure 2 shows the ground-state energy per particle $E_g$ as a function of inverse particle number 
$1/N$ for several valley-polarized states at $\nu = 1/2$, with spin and mixed polarizations fixed at $S = M = 0$. The finite-size energies exhibit a smooth and nearly linear dependence on $1/N$, allowing for a reliable extrapolation to the thermodynamic limit.

Among the states considered, the configuration labeled by flux-attachment parameters 
$(1 1 1 1 2)$ with zero valley polarization $(V=0)$ is found to have the lowest extrapolated ground-state energy. As the magnitude of valley polarization increases, the ground-state energy monotonically increases, indicating that valley-unpolarized or weakly polarized states are energetically favored at $\nu = 1/2$. Fully valley-polarized states $(V=±1)$ lie significantly higher in energy, suggesting that spontaneous full valley polarization is not favored in the absence of explicit symmetry-breaking terms.

This behavior reflects the multi-component nature of the SU(4) system, where distributing particles among different valley sectors allows for more efficient correlation effects and lowers the Coulomb energy. 
The small energy differences among several competing states indicate that the present results should be interpreted as identifying energetically competitive candidate states within the present variational framework rather than establishing a strict quantitative energy hierarchy. A more definitive distinction among nearly degenerate states would require additional finite-size and numerical uncertainty analyses, which lie beyond the scope of the present work.

\subsection{Filling fraction $\nu = 1/4$}
The ground-state energies for various polarized states at $\nu = 1/4$ are shown in Fig. 3. In this case, we consider both positively and negatively valley-polarized configurations with $S=M=0$, corresponding to different distributions of composite fermions among the four SU(4) components.

Unlike the $\nu = 1/2$ case, the energy differences between competing states at  $\nu = 1/4$ are relatively small, indicating a stronger competition between different polarization sectors. Nevertheless, a clear energetic hierarchy emerges, with partially valley-polarized states exhibiting lower energies compared to fully valley-polarized configurations. This suggests that Coulomb interactions favor states that retain some degree of SU(4) symmetry, even at this lower filling fraction.

The sensitivity of the ground-state energy to the choice of flux-attachment parameters highlights the importance of inter-component correlations in stabilizing candidate EDFQH states at  $\nu = 1/4$. 
The results suggest that multi-component correlated states constitute energetically competitive candidates for the observed $\nu=1/4$ state.

\subsection{Physical implications}

Taken together, our results suggest that even-denominator fractional quantum Hall states at 
$\nu = 1/2$ and $\nu = 1/4$ in monolayer graphene are stabilized by SU(4) multi-component correlations, with valley polarization playing a central role.  The energetically favored candidate states are neither fully valley polarized nor resemble a simple composite-fermion Fermi-sea description within the restricted family of trial states considered here.



Our calculations neglect effects such as gate screening, finite layer thickness, and Landau level mixing. These effects may influence the energetic ordering of competing states and therefore require further investigation. Incorporating such effects would be a natural direction for future work and could improve quantitative comparison with experiments.

\section{Discussion}
The results presented in this work provide insight into the nature of even-denominator fractional quantum Hall states in monolayer graphene within an SU(4) framework. By systematically comparing the ground-state energies of candidate states with different spin, valley, and mixed polarization configurations at $\nu = 1/2$ and $\nu = 1/4$, we find that certain multi-component SU(4) candidate states exhibit lower variational energies within the family of trial states considered. 

A key outcome of our analysis is the prominent role of valley degrees of freedom in determining the stability of EDFQH states. At $\nu = 1/2$, the lowest-energy state corresponds to a valley-unpolarized configuration, while increasing valley polarization leads to a monotonic increase in the ground-state energy. This suggests that, in the absence of explicit valley-symmetry-breaking fields, Coulomb interactions favor states that distribute particles among multiple valley components, thereby enhancing correlation effects. A similar tendency is observed at $\nu = 1/4$, although the competition between different polarization sectors is stronger, leading to smaller energy differences among candidate states.

The energetic preference for partially or unpolarized valley configurations highlights an important distinction between graphene and conventional semiconductor-based two-dimensional electron systems. In graphene, inter-valley scattering requires large momentum transfer and is therefore suppressed, allowing valley polarization to remain a soft degree of freedom compared to spin polarization. This asymmetry enables the stabilization of multi-component EDFQH states that are not naturally captured within single-component composite fermion descriptions. 
Our findings suggest that within the limited set of variational trial states considered in the present analysis, the energetically favored candidate states at $\nu = 1/2$ and $\nu = 1/4$ are more naturally described within the multi-component SU(4) Chern–Simons framework than by a simple composite-fermion Fermi-sea picture.

Instead, they are better described as correlated phases arising from nontrivial inter-component flux attachment in the SU(4) Chern–Simons framework. This interpretation is consistent with experimental observations \cite{Zibrov:2017, Zibrov:2018} of robust even-denominator states in high-quality monolayer graphene devices.

At ($\nu = 1/4$), the number of competing candidate configurations increases substantially due to the larger degeneracy of admissible multicomponent states. Consequently, we restrict our analysis to the fully valley-polarized sectors ($V=\pm 1$) and compare candidate states with different flux-attachment parameters. Interestingly, near-degenerate competing states emerge within these sectors, suggesting enhanced competition between multicomponent correlations at quarter filling. This behaviour contrasts with the clearer energetic ordering observed at ($\nu = 1/2$).


We note that the present work focuses specifically on multi-component SU(4) Halperin-like candidate states within a Chern–Simons composite-fermion framework. Other competing descriptions of even-denominator states in graphene have also been proposed, including Pfaffian, particle-hole-symmetric Pfaffian, and 221-parton states, particularly in higher Landau levels \cite{Balram:2015, Kim:2019}. Previous theoretical studies have suggested that these candidate states become energetically competitive depending on the Landau-level index, interaction details, and symmetry-breaking effects. A direct comparison with these competing topological orders lies beyond the scope of the present work.

Experimentally, even-denominator states at $\nu =\pm 1/2$ and $\nu=\pm 1/4$ were observed near an isospin transition in monolayer graphene \cite{Zibrov:2018}. The present Coulomb-only SU(4)-symmetric treatment does not explicitly include hBN-induced sub-lattice anisotropy, Zeeman coupling, or Landau-level mixing. Nevertheless, the identification of nearly competing multi-component states suggests that relatively weak symmetry-breaking fields may play an important role in selecting the experimentally realized phase \cite{MacDonald:2014, Kharitonov:2012, Zibrov:2018}.


\section{CONCLUSION}
In summary, we have investigated even-denominator fractional quantum Hall states in monolayer graphene at filling fractions $\nu = 1/2$ and $\nu = 1/4$ within an SU(4) Chern–Simons composite fermion framework. By constructing multi-component variational wave functions and evaluating their ground-state energies using the Coulomb interaction, we compared the stability of states with different valley polarization configurations.
 Our variational energy comparison suggests that the energetically favored candidate states are consistent with multi-component correlated quantum Hall liquids within the SU(4) framework.
These findings highlight the importance of SU(4) correlations in graphene and provide a unified theoretical perspective on the polarization structure of even-denominator fractional quantum Hall states.
\section*{Acknowledgements} 
One of the authors, MI thanks Saha Institute of Nuclear Physics for the post-doctoral fellowship fund.
\section*{Data availability statement} All data that support the findings of this study are included within the article.
\section*{ORCID iDs} 
\begin{itemize}
    \item{M. Indra: 0000-0002-7900-7947}
    \item{S. Sahu: 0000-0003-3220-8677}
  \end{itemize}
\bibliography{reference.bib} 

@article{Tsui_PRL:1982,
  title={Two-dimensional magnetotransport in the extreme quantum limit},
  author={Tsui, Daniel C and Stormer, Horst L and Gossard, Arthur C},
  journal={Physical Review Letters},
  volume={48},
  number={22},
  pages={1559},
  year={1982},
  doi={10.1103/PhysRevLett.48.1559},
  publisher={APS}
}

@article{Tsui_PRB:1982,
  title={Zero-resistance state of two-dimensional electrons in a quantizing magnetic field},
  author={Tsui, DC and St{\"o}rmer, HL and Gossard, AC},
  journal={Physical Review B},
  volume={25},
  number={2},
  pages={1405},
  year={1982},
  doi={10.1103/PhysRevB.25.1405},
  publisher={APS}
}

@article{Jain:1989,
  title={Composite-fermion approach for the fractional quantum Hall effect},
  author={Jain, Jainendra K},
  journal={Physical review letters},
  volume={63},
  number={2},
  pages={199},
  year={1989},
  doi={10.1103/PhysRevLett.63.199},
  publisher={APS}
}

@article{Dean:2011,
  title = {Multicomponent fractional quantum Hall effect in graphene},
  author = {Dean et. al.},
  journal = {Nature Physics},
  volume = {7},
  number = {9},
  pages = {693-696},
  year = {2011},
  doi = {10.1038/nphys2007},
  URL = {https://doi.org/10.1038/nphys2007}
}

@article{Modak:2011,
  title = {Fermionic Chern-Simons theory of SU(4) fractional quantum Hall effect},
  author = {Modak, Sanhita and Mandal, Sudhansu S. and Sengupta, K.},
  journal = {Phys. Rev. B},
  volume = {84},
  issue = {16},
  pages = {165118},
  numpages = {5},
  year = {2011},
  month = {Oct},
  publisher = {American Physical Society},
  doi = {10.1103/PhysRevB.84.165118},
  url = {https://link.aps.org/doi/10.1103/PhysRevB.84.165118}
}

@article{Lopez:1991,
  title={Fractional quantum Hall effect and Chern-Simons gauge theories},
  author={Lopez, Ana and Fradkin, Eduardo},
  journal={Physical Review B},
  volume={44},
  number={10},
  pages={5246},
  year={1991},
  doi={10.1103/PhysRevB.44.5246},
  publisher={APS}
}

@article{Balram:2015,
  title={Phase diagram of fractional quantum Hall effect of composite fermions in multicomponent systems},
  author={Balram, Ajit C and T{\H{o}}ke, Csaba and W{\'o}js, Arkadiusz and Jain, Jainendra K},
  journal={Physical review B},
  volume={91},
  number={4},
  pages={045109},
  year={2015},
  doi={10.1103/PhysRevB.91.045109}, 
  publisher={APS}
}

@article{Haldane:1983,
  title={Fractional quantization of the Hall effect: A hierarchy of incompressible quantum fluid states},
  author={Haldane, F Duncan M},
  journal={Physical Review Letters},
  volume={51},
  number={7},
  pages={605},
  year={1983},
  doi={10.1103/PhysRevLett.51.605},
  publisher={APS}
}

@article{Indra:2024_3,
doi = {10.1088/1402-4896/ad224f},
url = {https://doi.org/10.1088/1402-4896/ad224f},
year = {2024},
month = {feb},
publisher = {IOP Publishing},
volume = {99},
number = {3},
pages = {035915},
author = {Indra, Moumita and Majumder, Dwipesh},
title = {Study of polarization for even-denominator fractional quantum Hall states in SU(4) Graphene},
journal = {Physica Scripta}
}

@article{Li:2017,
author = {Li et. al.},
title = {Even-denominator fractional quantum Hall states in bilayer graphene},
journal = {Science},
volume = {358},
number = {6363},
pages = {648-652},
year = {2017},
doi = {10.1126/science.aao2521},
URL = {https://www.science.org/doi/abs/10.1126/science.aao2521},
eprint = {https://www.science.org/doi/pdf/10.1126/science.aao2521}}

@article{Ajit:2015,
  title = {Fractional quantum Hall effect in graphene: Quantitative comparison between theory and experiment},
  author = {Balram, Ajit C. and T\ifmmode \mbox{\H{o}}\else \H{o}\fi{}ke, Csaba and W\'ojs, A. and Jain, J. K.},
  journal = {Phys. Rev. B},
  volume = {92},
  issue = {7},
  pages = {075410},
  numpages = {13},
  year = {2015},
  month = {Aug},
  publisher = {American Physical Society},
  doi = {10.1103/PhysRevB.92.075410},
  url = {https://link.aps.org/doi/10.1103/PhysRevB.92.075410}
}

@article{Shabani:2009,
  title = {Evidence for Developing Fractional Quantum Hall States at Even Denominator $1/2$ and $1/4$ Fillings in Asymmetric Wide Quantum Wells},
  author = {Shabani, J. and Gokmen, T. and Chiu, Y. T. and Shayegan, M.},
  journal = {Phys. Rev. Lett.},
  volume = {103},
  issue = {25},
  pages = {256802},
  numpages = {4},
  year = {2009},
  month = {Dec},
  publisher = {American Physical Society},
  doi = {10.1103/PhysRevLett.103.256802},
  url = {https://link.aps.org/doi/10.1103/PhysRevLett.103.256802}
}

@article{Wang:2022,
  title = {Even-Denominator Fractional Quantum Hall State at Filling Factor $\ensuremath{\nu}=3/4$},
  author = {Wang et. al.},
  journal = {Phys. Rev. Lett.},
  volume = {129},
  issue = {15},
  pages = {156801},
  numpages = {6},
  year = {2022},
  month = {Oct},
  publisher = {American Physical Society},
  doi = {10.1103/PhysRevLett.129.156801},
  url = {https://link.aps.org/doi/10.1103/PhysRevLett.129.156801}
}

@article{Wang:2025,
  title = {Developing Fractional Quantum Hall States at Even-Denominator Fillings $1/6$ and $1/8$},
  author = {Wang et. al.},
  journal = {Phys. Rev. Lett.},
  volume = {134},
  issue = {4},
  pages = {046502},
  numpages = {7},
  year = {2025},
  month = {Jan},
  publisher = {American Physical Society},
  doi = {10.1103/PhysRevLett.134.046502},
  url = {https://link.aps.org/doi/10.1103/PhysRevLett.134.046502}
}

@article{Hossain:2018,
  title = {Unconventional Anisotropic Even-Denominator Fractional Quantum Hall State in a System with Mass Anisotropy},
  author = {Hossain, Md. Shafayat and Ma, Meng K. and Chung, Y. J. and Pfeiffer, L. N. and West, K. W. and Baldwin, K. W. and Shayegan, M.},
  journal = {Phys. Rev. Lett.},
  volume = {121},
  issue = {25},
  pages = {256601},
  numpages = {7},
  year = {2018},
  month = {Dec},
  publisher = {American Physical Society},
  doi = {10.1103/PhysRevLett.121.256601},
  url = {https://link.aps.org/doi/10.1103/PhysRevLett.121.256601}
}

@misc{sha:2025,
      title={Cascade of Even-Denominator Fractional Quantum Hall States in Mixed-Stacked Multilayer Graphene}, 
      author={Yating Sha et. al.},
      year={2025},
      eprint={2507.20695},
      archivePrefix={arXiv},
      primaryClass={cond-mat.mes-hall},
      url={https://arxiv.org/abs/2507.20695} 
}

@article{Huang:2026,
  title = {Non-Abelian fractional quantum Hall states at filling factor $3/4$},
  author = {Kai-Wen Huang et. al.},
  journal = {Chinese Physics B},
  volume = {Accepted Manuscrpt},
  issue = {},
  pages = {},
  numpages = {},
  year = {2026},
  month = {Jan},
  publisher = {IOP Publishing Ltd.},
  doi = {10.1088/1674-1056/ae3304},
  url = {https://iopscience.iop.org/article/10.1088/1674-1056/ae3304}
}

@article{Papic:2011,
  title = {Tunable Electron Interactions and Fractional Quantum Hall States in Graphene},
  author = {Papi\ifmmode \acute{c}\else \'{c}\fi{}, Z. and Thomale, R. and Abanin, D. A.},
  journal = {Phys. Rev. Lett.},
  volume = {107},
  issue = {17},
  pages = {176602},
  numpages = {5},
  year = {2011},
  month = {Oct},
  publisher = {American Physical Society},
  doi = {10.1103/PhysRevLett.107.176602},
  url = {https://link.aps.org/doi/10.1103/PhysRevLett.107.176602}
}

@article{Eisenstein:1994,
  title = {Compressibility of the two-dimensional electron gas: Measurements of the zero-field exchange energy and fractional quantum Hall gap},
  author = {Eisenstein, J. P. and Pfeiffer, L. N. and West, K. W.},
  journal = {Phys. Rev. B},
  volume = {50},
  issue = {3},
  pages = {1760--1778},
  numpages = {0},
  year = {1994},
  month = {Jul},
  publisher = {American Physical Society},
  doi = {10.1103/PhysRevB.50.1760},
  url = {https://link.aps.org/doi/10.1103/PhysRevB.50.1760}
}

@article{Young:2012,
  title = {Spin and valley quantum Hall ferromagnetism in graphene},
  author = {Young et. al.},
  journal = {Nature Physics},
  volume = {8},
  issue = {7},
  pages = {550-556},
  numpages = {6},
  year = {2012},
  month = {Jul},
  publisher = {Springer Nature},
  doi = {10.1038/nphys2307},
  url = {https://doi.org/10.1038/nphys2307}
}

@article{Sujit:2018,
  title = {Incompressible even denominator fractional quantum Hall states in the zeroth Landau level of monolayer graphene},
  author = {Narayanan, Sujit and Roy, Bitan and Kennett, Malcolm P.},
  journal = {Phys. Rev. B},
  volume = {98},
  issue = {23},
  pages = {235411},
  numpages = {5},
  year = {2018},
  month = {Dec},
  publisher = {American Physical Society},
  doi = {10.1103/PhysRevB.98.235411},
  url = {https://link.aps.org/doi/10.1103/PhysRevB.98.235411}
}

@article{Kim:2019,
 title = {Even denominator fractional quantum Hall states in higher Landau levels of graphene},
  author = {Kim et. al.},
  journal = {Nature Physics},
  volume = {15},
  issue = {2},
  pages = {154-158},
  numpages = {5},
  year = {2019},
  month = {Feb},
  publisher = {Springer Nature},
  doi = {10.1038/s41567-018-0355-x},
  url = {https://doi.org/10.1038/s41567-018-0355-x}
}

@article{Zibrov:2017,
  title = {Tunable interacting composite fermion phases in a half-filled bilayer-graphene Landau level},
  author = {Zibrov et. al.},
  journal = {Nature},
  volume = {549},
  issue = {7672},
  pages = {364-549},
  numpages = {6},
  year = {2017},
  month = {Sept},
  publisher = {Springer Nature},
  doi = {10.1038/nature23893},
  url = {https://doi.org/10.1038/nature23893}
}

@article{Zibrov:2018,
  title = {Even-denominator fractional quantum Hall states at an isospin transition in monolayer graphene},
  author = {Zibrov et. al.},
  journal = {Nature Physics},
  volume = {14},
  issue = {9},
  pages = {930-935},
  numpages = {5},
  year = {2018},
  month = {Jul},
  publisher = {Springer Nature},
  doi = {10.1038/s41567-018-0190-0},
  url = {https://doi.org/10.1038/s41567-018-0190-0}
}

@article{MacDonald:2014,
  title = {Broken SU(4) Symmetry and the Fractional Quantum Hall Effect in Graphene},
  author = {Sodemann, I. and MacDonald, A. H.},
  journal = {Phys. Rev. Lett.},
  volume = {112},
  issue = {12},
  pages = {126804},
  numpages = {5},
  year = {2014},
  month = {Mar},
  publisher = {American Physical Society},
  doi = {10.1103/PhysRevLett.112.126804},
  url = {https://link.aps.org/doi/10.1103/PhysRevLett.112.126804}
}

@article{Kharitonov:2012,
  title = {Phase diagram for the $\ensuremath{\nu}=0$ quantum Hall state in monolayer graphene},
  author = {Kharitonov, Maxim},
  journal = {Phys. Rev. B},
  volume = {85},
  issue = {15},
  pages = {155439},
  numpages = {23},
  year = {2012},
  month = {Apr},
  publisher = {American Physical Society},
  doi = {10.1103/PhysRevB.85.155439},
  url = {https://link.aps.org/doi/10.1103/PhysRevB.85.155439}
}

@article{Nayak:2008,
  title = {Non-Abelian anyons and topological quantum computation},
  author = {Nayak, Chetan and Simon, Steven H. and Stern, Ady and Freedman, Michael and Das Sarma, Sankar},
  journal = {Rev. Mod. Phys.},
  volume = {80},
  issue = {3},
  pages = {1083--1159},
  numpages = {0},
  year = {2008},
  month = {Sep},
  publisher = {American Physical Society},
  doi = {10.1103/RevModPhys.80.1083},
  url = {https://link.aps.org/doi/10.1103/RevModPhys.80.1083}
}

@article{Goerbig:2008,
  title = {Plasma picture of the fractional quantum Hall effect with internal $\text{SU}(K)$ symmetries},
  author = {de Gail, R. and Regnault, N. and Goerbig, M. O.},
  journal = {Phys. Rev. B},
  volume = {77},
  issue = {16},
  pages = {165310},
  numpages = {13},
  year = {2008},
  month = {Apr},
  publisher = {American Physical Society},
  doi = {10.1103/PhysRevB.77.165310},
  url = {https://link.aps.org/doi/10.1103/PhysRevB.77.165310}
}

@article{Halperin:1984,
  title = {Statistics of Quasiparticles and the Hierarchy of Fractional Quantized Hall States},
  author = {Halperin, B. I.},
  journal = {Phys. Rev. Lett.},
  volume = {52},
  issue = {18},
  pages = {1583--1586},
  numpages = {0},
  year = {1984},
  month = {Apr},
  publisher = {American Physical Society},
  doi = {10.1103/PhysRevLett.52.1583},
  url = {https://link.aps.org/doi/10.1103/PhysRevLett.52.1583}
}

@article{Geim:2009,
  title = {The electronic properties of graphene},
  author = {Castro Neto, A. H. and Guinea, F. and Peres, N. M. R. and Novoselov, K. S. and Geim, A. K.},
  journal = {Rev. Mod. Phys.},
  volume = {81},
  issue = {1},
  pages = {109--162},
  numpages = {0},
  year = {2009},
  month = {Jan},
  publisher = {American Physical Society},
  doi = {10.1103/RevModPhys.81.109},
  url = {https://link.aps.org/doi/10.1103/RevModPhys.81.109}
}

@article{MacDonald:2006,
  title = {Quantum Hall Ferromagnetism in Graphene},
  author = {Nomura, Kentaro and MacDonald, Allan H.},
  journal = {Phys. Rev. Lett.},
  volume = {96},
  issue = {25},
  pages = {256602},
  numpages = {4},
  year = {2006},
  month = {Jun},
  publisher = {American Physical Society},
  doi = {10.1103/PhysRevLett.96.256602},
  url = {https://link.aps.org/doi/10.1103/PhysRevLett.96.256602}
}

@article{Shibata:2008,
  title = {Coupled charge and valley excitations in graphene quantum Hall ferromagnets},
  author = {Shibata, Naokazu and Nomura, Kentaro},
  journal = {Phys. Rev. B},
  volume = {77},
  issue = {23},
  pages = {235426},
  numpages = {5},
  year = {2008},
  month = {Jun},
  publisher = {American Physical Society},
  doi = {10.1103/PhysRevB.77.235426},
  url = {https://link.aps.org/doi/10.1103/PhysRevB.77.235426}
}
\end{document}